\documentstyle[aps,psfig,multicol]{revtex}

\title{Suppression of synchrotron radiation due to beam crystallization}
\author{Harel Primack 
          \footnote{email: harel@phyc1.physik.uni-freiburg.de}
        and Reinhold Bl\"umel
          \footnote{email: blumel@phyc1.physik.uni-freiburg.de}
       }
\address{Fakult\"{a}t f\"{u}r Physik, Albert-Ludwigs-Universit\"{a}t,
         Hermann-Herder-Str.\ 3, D-79104 Freiburg, Germany}
\date{To appear in Eur. Phys. J. A, December 1998 issue}

\begin{document}
\maketitle

\begin{abstract}
%
With respect to a ``hot'', non-crystallized beam the synchrotron
radiation of a cold crystallized beam is considerably modified. We
predict suppression of synchrotron radiation emitted by a crystallized
beam in a storage ring. We also propose experiments to detect this
effect.
%
\end{abstract}

\vspace{0.2cm}

\begin{multicols}{2}
\narrowtext

At the turn of the century J.~J.~Thomson introduced the ``raisin
cake'' model for the structure of atoms \cite{Th1}. In this model, the
point--like electrons are immersed and circulate in a uniform
positively--charged background. One of the interesting features of
this model is the suppression of electromagnetic radiation emitted by
the circulating electrons due to destructive interference. Thomson
showed, for example, that six equi--spaced circulating electrons emit
approximately $10^{-17}$ of the power of a single electron \cite{Th2}
when travelling at 1\% of the speed of light. As a model for the
naturally occurring atoms of the periodic table Thomson's raisin cake
model was quickly replaced by the planetary models of Rutherford and
Bohr. However, Thomson's model is currently experiencing a revival in
the context of ``artificial atoms'' (see, e.g., charged particle traps
\cite{Bl}, quantum dots \cite{Dot} and clusters \cite{Haberland}). 
Moreover, present--day technology is very close to producing
space--charge dominated structures that will allow us to use (and
elaborate on) the ideas of Thomson.

In the context of accelerator physics, suppression of synchrotron
radiation by equi--spaced charges was discussed by Schiff \cite{Sch46}
in 1946. However, he remarks that ``\ldots it is difficult to see why
the equally spaced configuration should persist''.

In 1985 Schiffer and Kienle suggested the possibility of crystallized
ion beams \cite{SK85}, i.e.\ geometrically--ordered structures that
circulate in storage rings with speeds comparable to that of light.
The spatial order is effected by the balance between Coulomb repulsion
and focusing forces in the presence of strong cooling. To date there
is no clear--cut experimental evidence for a fast--moving ion--beam
crystal. Nevertheless, taking into account the efforts invested in
such projects \cite{HG95} as well as the advances in cooling
techniques \cite{HKNPS91,MGGHS96}, it seems likely that the goal of
producing a crystalline beam will be achieved in the foreseeable
future. In fact, considering the parameter regimes in which the
experimentalists work, one expects that the first crystalline beams
will be one--dimensional chains of equi-spaced ions, reminiscent of a
scaled-up Thomson atom. Thus, we can now offer a solution to Schiff's
remark in the form of crystallized beams.

In this note we propose the use of crystalline beams (especially 1D
chains) to achieve {\em suppression of synchrotron radiation} due to
beam crystallization. In the rest of this note we shall discuss these
issues and suggest experimental realizations.

The suppression of synchrotron radiation is based directly on the
ideas of Thomson described above. The importance of this effect is the
potential possibility to accelerate particles to very high energies
such that the synchrotron radiation does not limit the final energy,
even for moderately--sized accelerators. Currently, this mainly
applies to electrons, since their synchrotron radiation is about
$10^{13}$ larger than for protons with the same energy.

Suppose that we have $N$ particles with charge $q$ which circulate in
a plane (e.g. due to a constant magnetic field) with velocity $v$ and
radius of curvature $\rho$. The power emitted by a single particle is
given by \cite{Jac75,LL79}:
\begin{equation}
  P^{(1)} = \frac{q^2}{6 \pi \epsilon_0}
            \frac{\beta^4 \gamma^4 c}{\rho^2} ,
\end{equation}
where $\beta \equiv v/c$, $\gamma \equiv 1/\sqrt{1-\beta^2}$, and $c$
is the velocity of light. Due to the periodicity of the motion, one
can rewrite $P^{(1)}$ as:
\begin{equation}
  P^{(1)} = \sum_{n=1}^{\infty} P^{(1)}_{n} \; ,
\end{equation}
where $P^{(1)}_{n}$ is the contribution to the total power of
radiation with frequency $\omega_n = n v / \rho \equiv n
\omega$. It is given by \cite{Sch12,KL66,LL79}:
\begin{equation}
  P^{(1)}_{n} = \frac{q^2}{2 \pi \epsilon_0} 
    \frac{c \beta}{\gamma^2 \rho^2}
    \left[ \beta^2 \gamma^2 n J_{2n}^{\prime}(2 n \beta) {-}
           n^2 \!\!\! \int_{0}^{\beta} \!\!\! J_{2n}(2 n \xi) {\rm d}\xi 
    \right] .
  \label{eq:p1n}
\end{equation}
where $J_n$ are the ordinary Bessel functions. For $N$ randomly placed
particles, the total power emitted is simply $N$ times that of a
single particle:
\begin{equation}
  P^{(N)}_{\rm random} = N P^{(1)} =
  N \sum_{n=1}^{\infty} P^{(1)}_{n} \; .
\end{equation}
For $N$ particles that are equi--spaced along the circular trajectory
we have the following expression \cite{Jac75}:
\begin{equation}
  P^{(N)}_{\rm crystal} = N^2 \sum_{m=1}^{\infty} 
    P^{(1)}_{m {\cdot} N} \; .
\end{equation}
Thus, if $P^{(1)}_{n}$ is a rapidly decreasing function of $n$, we get
an overall {\em suppression of synchrotron radiation}. Examining
equation (\ref{eq:p1n}), this is indeed the case for all $\beta$ and
for large enough $N$. To show the suppression effect in detail we plot
the suppression factor $\alpha(N, \beta) \equiv P^{(N)}_{\rm crystal}
/ P^{(N)}_{\rm random}$ as a function of $N$ for $\beta \gamma = 0.1,
1, 10$ (sub-, intermediate and ultra-relativistic regimes).
\begin{figure}[t]

   \centerline{\psfig{figure=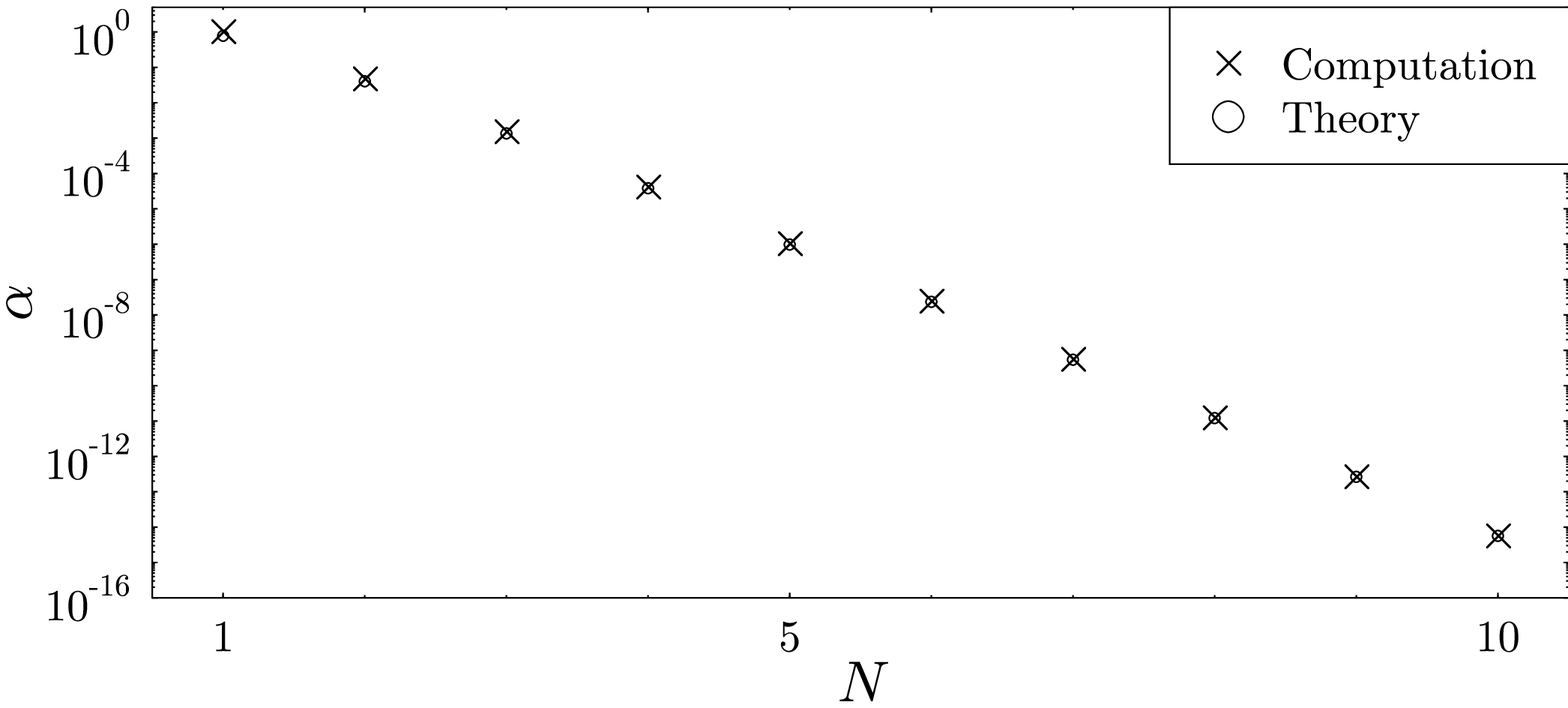,height=3.5cm}}
   \centerline{\psfig{figure=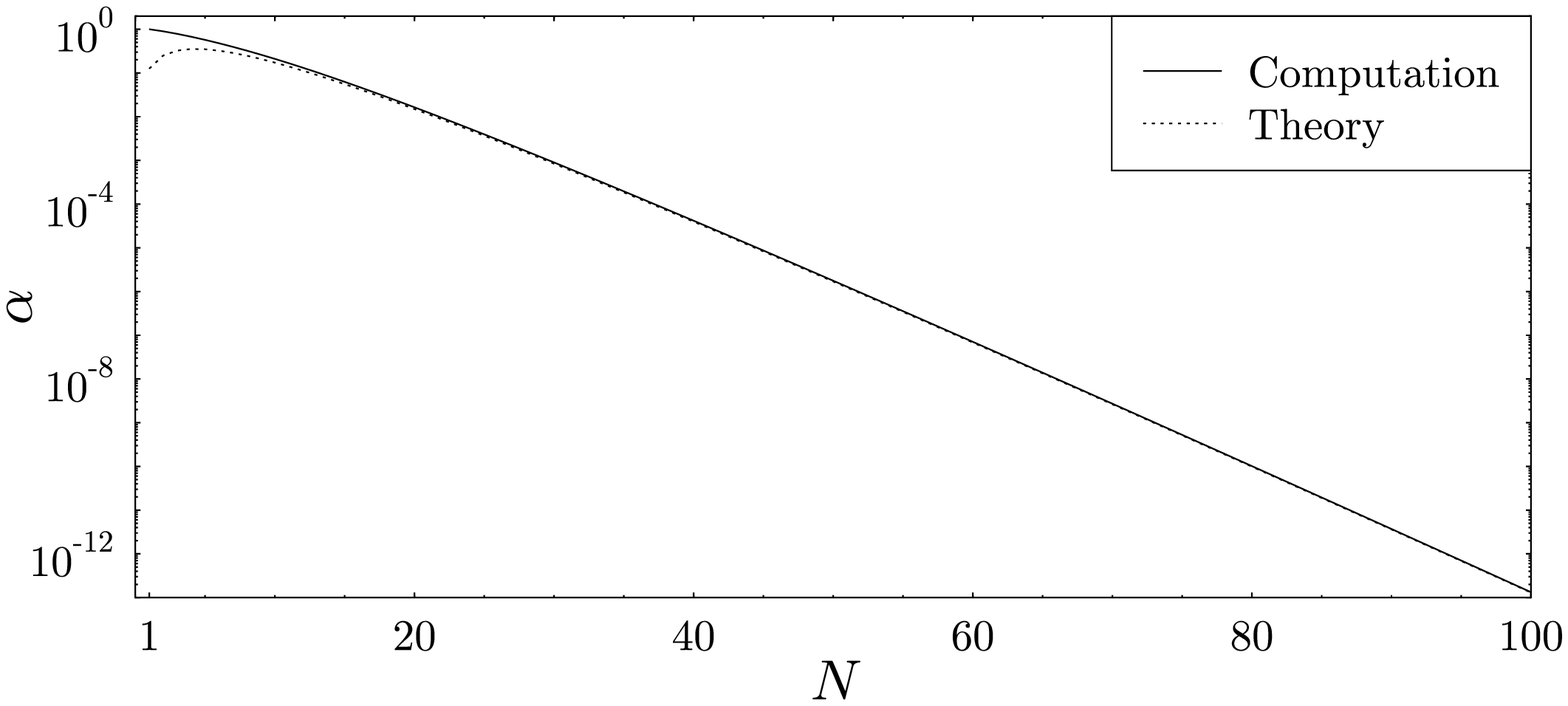,height=3.5cm}}
   \centerline{\psfig{figure=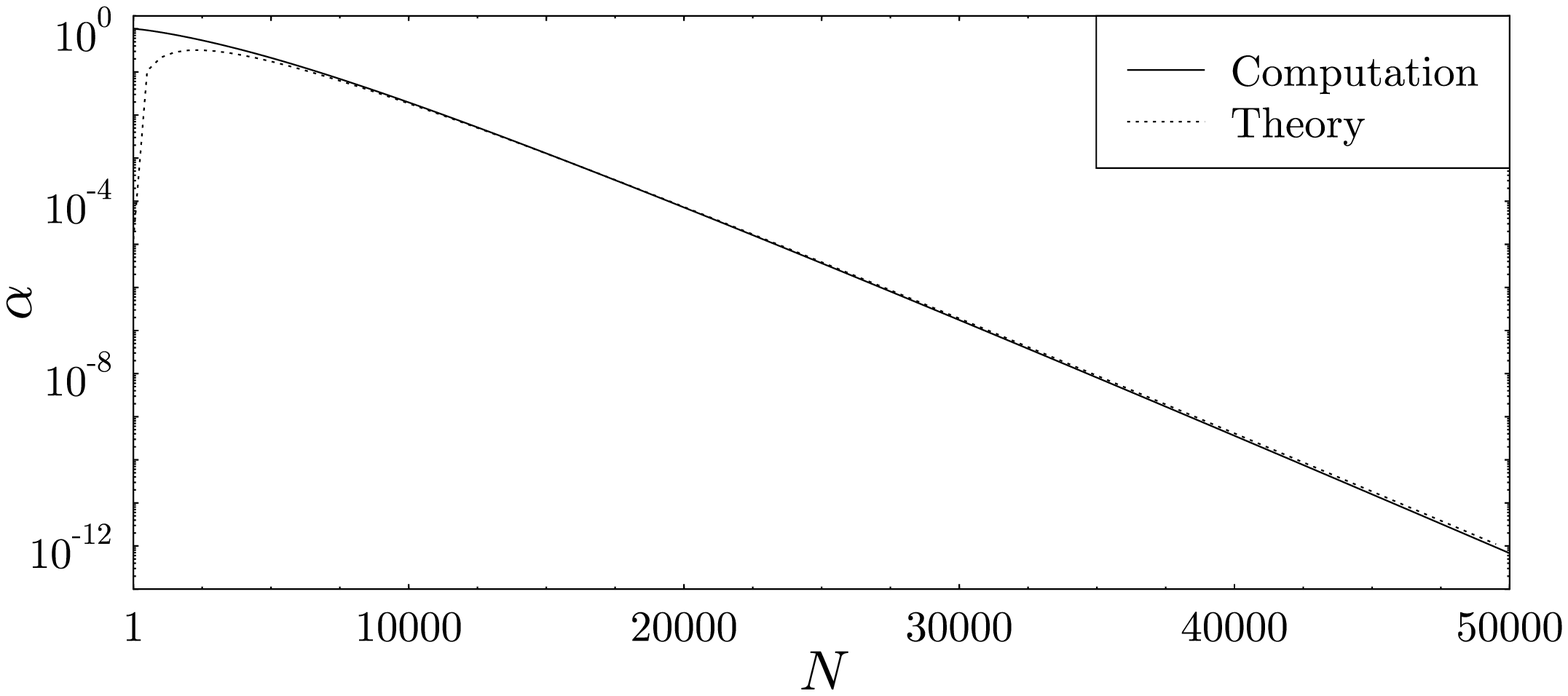,height=3.5cm}}
    \vspace{0.2cm}

  \caption{Suppression factor for equi--spaced circulating particles
     (crystallized beam). Upper plot: $\beta \gamma = 0.1$
     (sub-relativistic), middle: $\beta \gamma = 1$ (intermediate),
     lower: $\beta \gamma = 10$ (ultra-relativistic). The theoretical
     curves correspond to equations (\protect\ref{eq:sup-int},
     \protect\ref{eq:sup-ult}).}

  \label{fig:suppression}

\end{figure}
We observe in all cases exponential suppression of the synchrotron
radiation for sufficiently large $N$. For sub-relativistic velocities
the exponential decay is manifest practically for all $N > 1$. For
intermediate velocities the exponential decay starts at $N \approx
10$. For ultra-relativistic velocities, the exponential decay is
present for $N \gg \gamma^3 \approx 1000$ in our case. In terms of
analytical expressions, one can derive the following results
(e.g. \cite{Jac75,LL79}):
\begin{eqnarray}
  \alpha(N, \beta) 
  & \approx &
  \frac{3 N^{\frac{3}{2}}}{4 \sqrt{\pi} \beta^2 \gamma^{\frac{9}{2}}} 
  \left( \frac{\beta \gamma {\rm e}^{\frac{1}{\gamma}}}
              {1 + \gamma} \right)^{2N}
  \; , \; \; \gamma \gtrsim 1 , N \gg 1 \; ,
  \label{eq:sup-int}
  \\
  \alpha(N, \beta) 
  & \approx &
  \frac{3 N^{\frac{3}{2}}}{4 \sqrt{\pi} \beta^2 \gamma^{\frac{9}{2}}} 
  \exp \left( - \frac{2N}{3 \gamma^3} \right)
  \; , \; \; \gamma \gg 1 , N \gg \gamma^3 \; .
  \label{eq:sup-ult}
\end{eqnarray}
The analytical formulas clearly fit the numerical computations in
their range of validity.

At currently achievable magnetic fields even ultra-relativistic ion
beams emit very little synchrotron radiation power. Hence, the
experimental observation of the suppression effect for ions is very
difficult. However, for ultra-relativistic electrons the synchrotron
radiation power is appreciable. Thus, our theoretical calculations may
motivate experiments on crystallization of electron beams. In order to
experimentally observe the suppression effect, we suggest two types of
experiments:
\begin{enumerate}

  \item Start from cold electrons at rest which form a circular chain,
     and accelerate them such that the crystalline order is
     maintained. The initial crystallization may be realized by
     resistive cooling techniques \cite{Ghosh}.

  \item Accelerate electrons and apply sympathetic cooling by a fast
     crystallized ion beam. This is the inverse mechanism of electron
     cooling that is used currently in storage rings
     \cite{HG95}. Sufficiently strong cooling of the electron beam
     leads to its crystallization, and hence to dramatic suppression
     of synchrotron radiation.

\end{enumerate}
Further work is needed in order to evaluate the feasibility of the
suggested experiments.

HP acknowledges a MINERVA scholarship. RB acknowledges financial
support from the Deutsche Forschungsgemeinschaft (SFB 276).



\end{multicols}

\end{document}